\documentclass[12pt,twoside]{article}

\usepackage{amssymb}
\usepackage{wasysym}
\usepackage{amsthm}
\usepackage{graphicx}
\usepackage{pst-all}
\usepackage{multido}
\def\R{{\mathbb R}}
\def\N{{\mathbb N}}
\def\C{{\mathbb C}}

\newcommand\ds{\displaystyle\sum}

\begin{document}
\pagestyle{myheadings} \markboth{ H. Gottschalk,H. Ouerdiane and B.
Smii }{ White noise convolution calculus and
Feynman graphs } \thispagestyle{empty}

\title{Convolution calculus on white noise spaces and Feynman graph representation of generalized renormalization flows}
 \author{Hanno Gottschalk${}^\sharp$, Habib Ouerdiane ${}^\flat$ and Boubaker Smii${}^{\flat\sharp}$  }
\maketitle {\small

 \noindent ${}^\flat$: D\'epartement de Math\'ematiques, Universit\'e de Tunis El Manar

 \noindent  ${}^\sharp$: Institut f\"ur angewandte Mathematik, Rheinische Fridrich-Wilhelms-Universit\"at Bonn

\vspace{1cm}

\noindent {\bf Abstract :} In this note we outline some novel
connections between the following fields: 1) Convolution calculus
on white noise spaces 2) Pseudo-differential operators and L\'evy
processes on infinite dimensional spaces 3) Feynman graph
representations of convolution semigroups 4) generalized
renormalization group flows and 5) the thermodynamic limit of
particle systems.

\vspace{1cm}



\vspace{1cm}

\noindent Convolution semigroups on infinite dimensional spaces are the mathematical backbone of the Wilson-Polchinski formulation of renormalization group flows \cite{Sa}. There, a "time" (in this context one should say: scale) dependent infinite dimensional heat equation
\begin{equation}
\label{1eqa}
{\partial\over \partial t} e^{-V_t^{\rm eff}(\phi)}=\dot \Delta_{t,T_0} e^{-V_t^{\rm eff}(\phi)}~ ,~~ V_{T_0}^{\rm eff}(\phi)=V^{\rm initial}(\phi)
\end{equation}
governs the flow of the effective action $V^{\rm eff}_t$ between two scales $T,T_0$, where $T$ is the scale on which the system is observed and $T_0$ is the cut-off scale. The "time" dependent infinite dimensional laplacian $\dot \Delta_{t,T_0}={\partial\over\partial_t}\Delta_{t,T_0}$ is defined by
\begin{equation}
\label{2eqa}
\Delta_{t,T_0}=\int_{\R^d}\int_{\R^d}G_{t,T_0}(x,y){\delta\over\delta\phi(x)}{\delta\over\delta\phi(y)}\, dxdy
\end{equation}
with $G_{t,T_0}=G_t-G_{T_0}$ the covariance function of the Gaussian random field that has to be integrated out
 to intermediate between the random field with fundamental cut-off scale $T_0$ and covariance $G_{T_0}$ and
 the random field with cut off $t$ and covariance $G_t$ describing Gaussian fluctuations at some other scale $t\not=T_0$ ($t>T_0$ for ultra violet and $t<T_0$ for infra red problems).

The functional derivative ${\delta\over\delta\phi(x)}$ acts on a function $F(\phi)=\sum_{n=0}^\infty\langle F_n,\phi^{\hat\otimes n}\rangle$ via ${\delta\over\delta\phi(x)}F(\phi)=\sum_{n=0}^\infty n\langle D_xF_n,\phi^{\hat \otimes n}\rangle$, $D_xF_n(x_2,\ldots,x_n)=F_n(x,x_2,\ldots,x_n)$. Here the kernel functions $F_n$ are assumed to be symmetric.

The usefulness and mathematical beauty of the Wilson-Polchinski approach to renormalization is given by the following key features
\begin{enumerate}
\item The renormalization equation is the generating equation of a infinite dimensional diffusion process;
\item The perturbative solution of \ref{1eqa}) can be represented as a sum over Feynman graphs.
\item Renormalization conditions at a scale $T$ can be imposed by a change of the initial condition $V^{\rm initial}$ at scale $T_0$ leading to a finite theory if the cut off $T_0$ is removed;
\end{enumerate}

\noindent In this note outline the generalization of (\ref{1eqa})
replacing the infinite dimensional Laplacian by a
pseudo-differential operator in infinite dimensions. We show that
under this generalization the key features of the renormalization
group approach all prevail: Instead of infinite dimensional
diffusion processes one obtains jump-diffusion type L\'evy
processes and instead of classical Feynman graphs generalized
Feynman graphs studied in \cite{DGO}. The idea of renormalization
as given in item 3 above then carries over unchanged. We will
illustrate this for the special case of particle systems in the
continuum.

White noise analysis is a natural framework to rigorously formulate equations like (\ref{1eqa}) and their generalizations:
In fact, following \cite{GHOR}, we let ${\cal S}={\cal S}(\R^d,\C)$ be the space of rapidly decreasing test functions equipped with the Schwartz topology which is generated by an increasing sequence of Hilbert seminorms $\{|.|_p\}_{p\in\N}$ and let ${\cal S}'$ the dual of ${\cal S}$.  By ${\cal S}_p$ we denote the closure of ${\cal S}$ wrt $|.|_p$ and by ${\cal S}_{-p}$ the topological dual of ${\cal S}_p$.
For $\theta(t)$, $t\leq0$, a Young function (nonnegative, continuous, convex and strictly increasing s.t. $\lim_{t\to\infty}\theta(t)/t=\infty$) we set $\theta^*(t)=\sup_{x\geq 0}(xt-\theta(x))$, the Legendre transform of $\theta$, which is another Young function.

Given a complex Banach space $(B, \parallel.
\parallel)$, let $H(B)$ denote the space of entire function on $B$, i.e the space of continuous functions from $B$ to $\C$, whose restriction
to all affine lines of $B$ are entire on $\C$. Let $Exp(B\,,
\theta\,, m)$ denote the space of all entire functions on $B$ with
exponential growth of order $\theta$, and of finite type $m>0$
\begin{equation}
\label{3eqa}
Exp(B,\theta,m)=\{f \in H(B)\,;\,
\|f\|_{\theta,m}=\sup_{z \in
B}|f(z)|e^{-\theta(m\|z\|)}<+\infty  \}
\end{equation}
In the following we consider the white noise test functions space
\begin{equation}
\label{4eqa}
 {\cal F}_{\theta}({\cal S}')=\bigcap_{p \geq 0\,, m>0}
Exp({\cal S}_{-p}\,, \theta\,, m).
\end{equation}
Let ${\cal F}_{\theta}({\cal S}')^*$, the space of white noise distributions, be the strong dual of the space ${\cal
F}_{\theta}(\cal S')$ equipped with the projective limit topology. For any $f\in{\cal S}'$ and $\theta$, the exponential $e^f:{\cal S}'\to\C$, $e^f(\phi)=e^{\langle\phi,f\rangle}$ is in ${\cal F}_\theta({\cal S}')$. Thus, the
Laplace transform ${\cal L}(\Phi)(f)=\langle\Phi,e^f\rangle$ is well defined for $f\in{\cal S}$. Recalling the definition of the space
\begin{equation}
\label{5eqa}
{\cal G}_\theta({\cal S})=\bigcup_{p\geq 0,m>0} Exp(N_p,\theta,m)
\end{equation}
which is equipped with the topology of the inductive limit, we get \cite{GHOR} that ${\cal L}:{\cal F}_\theta({\cal S}')^*\to {\cal G}_{\theta^*}({\cal S})$ is a topological isomorphism. Using the property of Young functions $\lim_{t\to\infty}\theta^*(t)/t=\infty$ it is easy to see that ${\cal G}_{\theta^*}$ is an algebra under multiplication.
Thus, for $\Psi,\Phi\in{\cal F}_\theta({\cal S}')^*$ one can define the convolution $\Psi*\Phi={\cal L}^{-1}({\cal L}(\Psi){\cal L}(\Phi))$ as an element of ${\cal F}_\theta({\cal S}')^*$.

Let us assume for a moment that $\lim_{t\to\infty}\phi(t)/t^2$
exists and is finite. Under this condition we have that ${\cal
F}_\theta({\cal S}')\to L^2_\C({\cal S}_\R',d\nu_0)\to{\cal
F}_\theta({\cal S}')^*$ is a Gelfand triplet, where $\nu_0$ is the
white noise measure, cf. \cite{GHOR}. Suppose that $\Psi\in{\cal
F_\theta}({\cal S'})$ has a Taylor series
$\Psi(\phi)=\sum_{n=0}^N\langle\Phi_n,\phi^{\hat \otimes
n}\rangle$ with $\Phi_n\in{\cal S}^{\hat\otimes n}$. One can then
show that $\Psi*\Phi\in{\cal F}_\theta({\cal S}')$ and
$\Psi*\Phi(\phi)=\Psi({\delta\over\delta\phi})\Phi(\phi)$ with
\begin{equation}
\label{6eqa}
\Psi({\delta\over\delta\phi})=\sum_{n= 0}^\infty\int_{\R^{dn}}\Psi_n(x_1,\ldots,x_n){\delta\over\delta\phi(x_1)}\cdots{\delta\over\delta\phi(x_n)}\,dx_1\cdots dx_n
\end{equation}
Here the rule for the evaluation of the pseudo-differential
operator $\Psi({\delta\over\delta\phi})$ is that first the $n$-th
order differential operators are applied to $\Phi$ and then the
result assumed up over all $n$. Hence we see that the equation
\begin{equation}
\label{7eqa}
{\partial\over \partial t} e^{-V_t^{\rm eff}}=\dot \Psi_{t,T_0}({\delta\over\delta\phi}) e^{-V_t^{\rm eff}}=\dot \Psi_{t,T_0}* e^{-V_t^{\rm eff}},~~ V_{T_0}^{\rm eff}=V^{\rm initial}
\end{equation}
is the correct generalization of the renormalization flow equation (\ref{1eqa}).
If now $\dot \Psi_{t,T_0}:\R_+\to{\cal F}_\theta({\cal S}')^*$ is continuous and
$e^{-V^{\rm initial}}\in{\cal F}_\theta({\cal S}')^*$, it has been proven in \cite{OP} that (\ref{7eqa}) has a unique solution in ${\cal F}_{(e^{\theta^*}-1)*}({\cal S}')^*$, namely
\begin{equation}
\label{8eqa}
e^{-V^{\rm eff}_t}={\cal L}^{-1}(e^{\int_{T_0}^t{\cal L}(\dot \Psi_{s,T_0})\, ds})*e^{-V^{\rm initial}}.
\end{equation}
Again, the above solution is of particular interest if a
probabilistic interpretation can be given. This is the case when
${\cal L}(\dot\Psi_{t,T_0})$ for every $t$ is a conditionally
positive function, i.e. ${\cal L}(\Psi_{t,T_0})(0)=0$,
\begin{equation}
\label{9eqa}
\sum_{j,l=1}^n{\cal L}(\dot \Psi_{t,T_0})(f_j+f_l)z_j\bar z_l\leq 0~~\forall t<T_0,n\in\N,f_l\in{\cal S}_\R,z_l\in\C:~\sum_{l=1}^nz_l=0
\end{equation}
and the opposite inequality holds for $t>T_0$.  Under these conditions, by the Bochner-Minlos theorem, the transition kernel ${\cal L}^{-1}(e^{\int_{T_0}^t{\cal L}(\dot \Psi_{t,T_0})\, dt})$ is a family of probability measures on ${\cal S}'$ that fulfills the Chapman-Kolmogorov equations and thus defines a stochastic process with state space ${\cal S}'$. In general, this process will be of jump-diffusion type, as it follows from the
L\'evy-It\^o decomposition of conditionally positive definite functions.

Let us now come to the Feynman graph expansion. We take an initial condition of the type
\begin{equation}
\label{10eqa}
V^{\rm initial}(\phi)=\ds_{p=0}^{\bar
p}\langle \lambda^{(p)},\phi^{\otimes p}\rangle
\end{equation}
with kernel (vertex) functions $\lambda^{(p)}\in{\cal S}^{\hat\otimes p}$, for simplicity (this condition can clearly be relaxed). It is also assumed that $V^{\rm initial}(\phi)\geq -C$ for some $C>0$ and all $\phi\in{\cal S}'$.
Then, $e^{-V^{\rm initial}}\in L^2({\cal S}',d\nu_0)$. If now $\theta(t)$ fulfills $\lim_{t\to\infty}\theta(t)/t^2=c<\infty$, then by the theorem cited above, the solution of (\ref{7eqa}) exists.

The next step is to expand (\ref{7eqa}) in a formal power series
in $V^{\rm initial}$. At least in the case where, for $t$ fixed,
${\cal L}^{-1}(e^{\int_{T_0}^t{\cal L}(\dot \Psi_{t,T_0})\, dt})$
is a measure on ${\cal S}_{\R}'$, this expansion is an asymptotic
series, cf. \cite[Lemma 2.2]{DGO}. We note that the Laplace
transform of a white noise distribution is an analytic function
\cite{GHOR}. One can thus consider the Taylor series in $f$ of
$e^{\int_{T_0}^t{\cal L}(\dot \Psi_{s,T_0})(f)\, ds}$ at zero
given by $$\sum_{m\geq 0}{1\over m!} \langle m_{n,t,T_0},f^{\hat
\otimes n}\rangle$$ for $f\in{\cal S}$ with $m_{n,t,T_0}\in({\cal
S}')^{\hat \otimes n}$ the $n$-th moment.  The connected moment
functions, $m_{n,t,T_0}^c$ by definition are the Taylor
coefficients of the logarithm of the generating functional of the
moment functions, i.e. , $$\int_{T_0}^t{\cal L}(\dot
\Psi_{s,T_0})(f)\, ds=\sum_{m\geq 0}{1\over m!} \langle
m^c_{n,t,T_0},f^{\hat \otimes n}\rangle$$ The well-known linked
cluster theorem, cf. e.g. \cite[Appendix A]{DGO}, then gives the
combinatorial relation between moments and connected moments,
namely
\begin{equation}
\label{11eqa}
m_{n,t,T_0}(x_1,\ldots,x_n)=\sum_{I\in{\cal P}(1,\ldots,n)\atop I=\{I_1,\ldots,I_k\}}\prod_{l=1}^km^c_{n_l,t,T_o}(x_{j_l^1},\ldots,x_{j_l^{n_l}})
\end{equation}
where ${\cal P}(1,\ldots,n)$ is the set of all partitions of
$\{1,\ldots,n\}$ into disjoint non empty subsets $I_1,\ldots,I_k$,
$k\in\N$ arbitrary, $I_l=\{j_l^1,\ldots,j_l^{n_l}\}$. After these
preparations one obtains by straight forward calculation
\begin{eqnarray} \label{12eqa}
e^{V^{\rm eff}_t(\phi)} &=& \sum_{m=0}^\infty
{{(-1)^m}\over {m!}}{\cal L}^{-1}(e^{\int_{T_0}^t{\cal L}(\dot \Psi_{s,T_0})\, ds})*(V^{\rm initial})^m (\phi)
\nonumber\\&=&\sum_{m=0}^\infty
{{(-1)^m}\over {m!}} \sum_{p_1,\ldots,p_m=0}^{\bar{p}}
\int_{\R^{d\sharp \Omega}}\lambda^{(p_1)} (x^1_1\cdots x^1_{p_1})\cdots\lambda^{(p_m)}(x^m_1,\cdots, x^m_{p_m})
\nonumber\\&\times&
\sum_{K \subset \Omega} \prod_{(s,\,q) \in K} (-\phi(x^q_s))\,
m_{t,T_0} (\Omega\setminus K)\otimes_{(s,q)\in \Omega}dx_s^q
\nonumber\\&=&\sum_{m=0}^\infty
{{(-1)^m}\over {m!}} \sum_{p_1,\ldots,p_m=0}^{\bar{p}}
\int_{\R^{d\sharp \Omega}}\lambda^{(p_1)} (x^1_1\cdots x^1_{p_1})\cdots\lambda^{(p_m)}(x^m_1,\cdots, x^m_{p_m})
\nonumber\\&\times&
\sum_{K \subset \Omega} \prod_{(s,\,q) \in K} (-\phi(x^q_s))
\sum_{{I\in{\cal P}(\Omega \setminus K)}\atop
I=\{I_1,\ldots,I_k\}}\prod_{l=1}^k m_{t,T_0}^c(I_l)\,\otimes_{(s,q)\in\Omega} dx_s^q.
\end{eqnarray}
Here we used the following notation:  $\Omega=\Omega(p_1,\ldots,p_m)=\{(1,1),\ldots,(1,p_1),\ldots,(m,1),\ldots,\linebreak(m,p_m)\}$ and for $A\subseteq \Omega$,
$A=\{(s_1,q_1),\ldots,(s_r,q_r)\}$, $m^{(c)}_{t,T_0}(A)=m_{r,t,T_0}^{(c)}(x_{s_1}^{q_1},\ldots,x_{s_r}^{q_r})$. ${\cal P}(\Omega\setminus K)$ is the set of partitions of $\Omega\setminus K$. We also made use of the fact that $(V^{\rm initial})^m(\phi)$, being a polynomial with test functions as coefficients,
 is a white noise test function and that the convolution between a white noise distribution $\Phi$ and a white noise test function $\Theta$ is $\Phi*\Theta(\phi)=\phi(\Theta_\phi)$ where $\Theta_\phi(\varphi)=\Theta(\varphi-\phi)$ is a shift.

Generalized Feynman graphs can now be used to order the combinatorial sum on the rhs of (\ref{12eqa}). A generalized (amputated) Feynman graph is a
graph with three types of vertices, called inner full $\bullet$, inner empty $\circ$ and
outer empty $\otimes$ vertices, respectively. By
definition full vertices are distinguishable and have
distinguishable legs whereas empty vertices are non distinguishable
and have non distinguishable legs. Outer empty vertices
are met by one edge only. Edges are non directed and connect full
and empty (inner and outer) vertices, but never connect two full or
two empty vertices. The set of generalized Feynman graphs with $m$
inner full vertices with $p_1,...,p_m$ the number of legs of the inner full vertices
 such that $p_j\leq\bar p$ and $\lambda^{(p_j)}\not=0$, $j=1,\ldots,m$, is denoted by $\bar F(m)$.
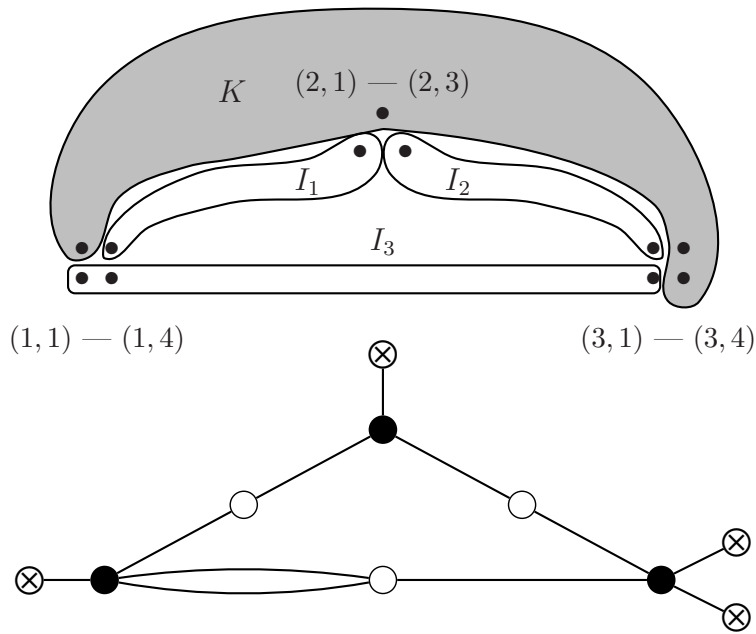
\begin{figure}
\centerline{\psset{xunit=1cm,yunit=1cm,runit=1cm,shortput=tab}
\begin{pspicture}(0,0)(10,8)
\pscurve*[linecolor=lightgray](5,7)(3,6.6)(2.4,6.5)(1.5,6.1)(1.1,5.3)(.8,5.3)(1.2,7.5)(5,8.6)(8.6,7.5)(9.2,4.7)(8.8,4.7)(8.8,5.5)(8.6,6)(5,7)
\pscurve(5,7)(3,6.6)(2.4,6.5)(1.5,6.1)(1.1,5.3)(.8,5.3)(1.2,7.5)(5,8.6)(8.6,7.5)(9.2,4.7)(8.8,4.7)(8.8,5.5)(8.6,6)(5,7)
\psccurve(4.9,6.9)(4.9,6.4)(2.9,6)(1.9,5.7)(1.3,5.28)(1.28,5.35)(1.4,5.7)(2.5,6.4)(4,6.6)(4.9,6.9)
\psccurve(5.1,6.9)(5.1,6.4)(7.1,6)(8.1,5.7)(8.7,5.28)(8.72,5.35)(8.6,5.7)(7.5,6.4)(6,6.6)(5.1,6.9)
\psframe[framearc=0.5](.8,4.8)(8.7,5.2)
\rput(5,7.2){$\bullet$}
\rput(4.7,6.7){$\bullet$}
\rput(5.3,6.7){$\bullet$}
\rput(5,7.6){\small $(2,1)$ --- $(2,3)$}
\rput(1.4,5){$\bullet$}
\rput(1,5){$\bullet$}
\rput(1,5.4){$\bullet$}
\rput(1.4,5.4){$\bullet$}
\rput(1.2,4.2){\small $(1,1)$ --- $(1,4)$}
\rput(8.6,5){$\bullet$}
\rput(9,5){$\bullet$}
\rput(9,5.4){$\bullet$}
\rput(8.6,5.4){$\bullet$}
\rput(8.8,4.2){\small $(3,1)$ --- $(3,4)$}
\rput(4,6.3){$I_1$}
\rput(6,6.3){$I_2$}
\rput(5,5.5){$I_3$}
\rput(3,7.5){$K$}
\dotnode[dotscale=3,dotstyle=*](5,3){A}
\dotnode[dotscale=3,dotstyle=*](1.3,1){B}
\dotnode[dotscale=3,dotstyle=*](8.7,1){C}
\dotnode[dotscale=3,dotstyle=otimes](5,4){D}
\dotnode[dotscale=3,dotstyle=otimes](0.3,1){E}
\dotnode[dotscale=3,dotstyle=otimes](9.7,1.5){H}
\dotnode[dotscale=3,dotstyle=otimes](9.7,0.5){I}
\dotnode[dotscale=3,dotstyle=o](5,1){J}
\dotnode[dotscale=3,dotstyle=o](3.15,2){K}
\dotnode[dotscale=3,dotstyle=o](6.85,2){L}
\ncline{A}{D}
\ncline{B}{E}
\ncline{C}{H}
\ncline{C}{I}
\ncline{A}{K}
\ncline[arcangle=10]{A}{L}
\ncarc[arcangle=-10]{B}{J}
\ncarc{B}{J}
\ncline{C}{J}
\ncline{B}{K}
\ncline{C}{L}
\end{pspicture}}
\caption{\label{4.1fig} Construction of a generalized Feynman graph from the set $K$ and the partition $I=\{I_1,I_2,I_3\}$}
\end{figure}

 To obtain the connection with (\ref{12eqa}) we consider an example where $m=3$, $p_1=4,p_2=3$ and $p_3=4$. For each element in $\Omega=\Omega(4,3,4)$ we draw a point s.t. points belonging to the same $p_l$ are
drawn close together. Then we choose a subset $K$ and a partition
$I$, cf. Fig. 1 (top). The generalized Feynman graph can now be
obtained by representing each collection of points by a full inner
vertex with $p_l$ legs, for each set $I_l$ in the partition we
draw a inner empty vertex connected to the legs of the inner full
vertices corresponding to the points in $I_l$. Finally we draw an
outer empty vertex connected to the leg of an inner full vertex
corresponding to each point in $K$. We then obtain the generalized
Feynman graph Fig. 1 (bottom). In this way, for fixed $m$, one
obtains a one to one correspondence between the index set of the
sum in (\ref{12eqa}) and ${\bar F}(m)$.

We want to calculate the contribution to (\ref{12eqa}) directly
from the graph $G\in\bar F(m)$ without the detour through the
above one to one correspondence. This is accomplished by the
following Feynman rules: Attribute an vector $x\in\R^d$ to each
leg of a inner full vertex. For each inner full vertex with $p$
legs multiply with $\lambda^{(p)}$ evaluated at the vectors
attributed to the legs of that vertex. For any inner empty vertex
with $l$ legs, multiply with a connected moment function
$m^{c}_{l,t,T_0}$ evaluated with the $l$ arguments corresponding
to the legs that this inner empty vertex is connected with. For
each outer empty vertex multiply with $-\phi(x)$ where $x$ is the
argument of the leg of the inner full vertex that the outer empty
vertex is connected with. Finally integrate $\int_{\R^d}\cdots dx$
over all the arguments $x$ that have been used to label the legs
of the inner full vertices. In this way one obtains the analytic
value ${\cal V}[G](t,T_0,\phi)$. The perturbative solution of
(\ref{7eqa}) then takes the form
\begin{equation}
\label{13eqa}
e^{-V_t^{\rm eff}(\phi)}=\sum_{m=0}^\infty{(-1)^m\over m!}\sum_{G\in\bar F(m)}{\cal V}[G](t,T_0,\phi),
\end{equation}
where the identity is in the sense of formal power series. The linked cluster theorem for generalized Feynman graphs proven in \cite{DGO,GS} then implies that $V^{\rm eff}_t$ can be calculated as a sum over connected Feynman graphs
\begin{equation}
\label{14eqa}
-V_t^{\rm eff}(\phi)=\sum_{m=0}^\infty{(-1)^m\over m!}\sum_{G\in\bar F_c(m)}{\cal V}[G](t,T_0,\phi).
\end{equation}
Let us now apply the above renormalization group scheme to the problem of taking the thermodynamic limit of a particle system. To this aim let $T=0$ and
\begin{equation}
\label{15eqa}
{\cal L}(\mu_{t})(f)=e^{\int_{\R^d}\int_{[-c,c]}\left(e^{sf(x)}-1 \right)dr(s)\sigma_{t}(x)dx}
\end{equation}
where $r$ is a probability measure on $[-c,c]$, $0<c<\infty$,
$\sigma_t(x)=\sigma(x/t)$ where $\sigma$ is a continuously
differentiable function with support in the unit ball and
$\nabla\sigma(0)=0$ such that $\sigma(0)=z>0$. ${\cal L}(\mu_{t})$
at the same time is the Laplace transform of the Poisson measure
$\mu_{t}$ representing a system of non-interacting, charged
particles in the grand canonical ensemble with intensity measure
(local density) $\sigma_{t}$, see e.g. \cite{AGY}.
 and a white noise distribution $\mu_{t}\in{\cal F}_{\theta}({\cal S}')^*$ for any $\theta$
 (due to the linear exponential growth of (\ref{15eqa}) in $f$ we have that the rhs is in ${\cal G}_{\theta^*}({\cal S})$ for any $\theta$), cf. \cite{GHOR}.
Both objects can thus be identified.
 Furthermore we assume that $\sigma (\alpha x)$ is monotonically decreasing in $\alpha$ for $\alpha>0$. Then $\dot \sigma_t(x)=d\sigma_t(x)/dt\geq0$ for $t>0,x\in\R^d$.  It is then standard to show that
\begin{equation}
\label{16eqa}
{\cal L}(\dot\Psi_t)(f)=-\int_{\R^d}\int_{[-c,c]}\left(e^{sf(x)}-1 \right)dr(s)\dot\sigma_t(x)dx
\end{equation}
in fact fulfills (\ref{9eqa}) for all $t<T_0$.  Thus the pseudo differential operator $\dot \Psi({\delta\over\delta\phi})$ is the generator of a jump-diffusion process with state space ${\cal S}'$ with backward time direction.
Let $\lambda^{(p)}=\lambda^{(p)}(x_1,\ldots,x_p)$ be a set of $C^{\infty}$ functions that are of rapid decrease in the difference variables $x_i-x_j$, $i\not=j$. For a distribution $\phi$ of compact support we define $V^{\rm initial}(\phi)$ as in (\ref{10eqa}). We assume that $V^{\rm initial}(\phi)>-C$ for each such $\phi$.

Let $\phi\in{\cal S}'$ have compact support. The non normalized correlation functional $\rho_t(\phi)$ of the particle system with infra-red cut-off $t$ is defined as
\begin{equation}
\label{17eqa}
\rho_t(\phi)=\int_{{\cal S}'}e^{-V^{\rm initial}(\varphi+\phi)}d\mu_t(\varphi)=\mu_t*e^{-V^{\rm initial}}(\phi).
\end{equation}
Note that $\mu_{T_0}$ has support on the distributions supported on a
ball of diameter $T_0$, $B_{T_0}$. Therefore, for $\phi$ fixed and $t<T_0$
one can replace $\lambda^{p}(x_1,\ldots,x_n)$ with
$\lambda^{(p)}(x_1,\ldots,x_n)\prod_{l=1}^n\chi(x_l)$ with $\chi $
being a test function that is constantly one on ${\rm
supp}\phi\cup B_{T_0}$ without changing (\ref{17eqa}). Under this
replacement, $V^{\rm initial}$ meets the conditions from above
that $\lambda^{(p)}\in{\cal S}^{\hat\otimes p}$.  
Furthermore, under this replacement in $V^{\rm initial}$, $e^{-V^{\rm initial}}\in L^2({\cal S}',d\nu_0)$ 
and hence $e^{-V^{\rm initial}}\in{\cal F}_\theta({\cal S}')^*$ if $\lim_{t\to\infty}\theta(t)/t^2$ finite.  Furthermore, one can argue as above to see that $\dot \Psi_t$ is in ${\cal F}_\theta({\cal S}')^*$ for $\theta$ arbitrary. Thereby, the requirements of white noise convolution calculus are met. We shall neglect
the inessential distinction between $V^{\rm initial}$ and its replaced version in the following.

From (\ref{15eqa}--\ref{17eqa}) we see that the non normalized correlation functional $\rho_t$ fulfills the renormalization group equation (\ref{7eqa}). The thermodynamic limit, which is achieved as $T_0\to\infty$, is thus governed by this equation (and thus by a L\'evy process with infinitely dimensional state space).

Let us now come to the issue of the normalization of $\rho_T$ at a time $T=0$, for simplicity. A normalized correlation functional should fulfill $\rho_T(0)=1$. But $-V_T^{\rm eff}(0)=\log\rho_T(0)\sim T_0^d$ in our case where the divergent (as $T_0\to\infty$) parts originate from the so-called vacuum to vacuum diagrams, i.e. such Feynman graphs in $\bar F_c(m)$ that do not have
outer empty vertices \cite[Thm 6.6]{DGO}. All other contributions remain finite in the limit $T_0\to\infty$ \cite[Section 7]{DGO}. The normalization $-V_T^{\rm eff}(0)=0$ can now be achieved perturbatively by re-defining $V^{\rm initial}$ by a counter term
\begin{equation}
\label{18eqa}
\delta\lambda^{(0)}_{T_0}=\sum_{m\geq0}{(-1)^m\over m!}\sum_{G\in \bar F_c^{\rm vac.\, to\, vac.}(m)}{\cal V}[G](0,T_0,0)
\end{equation}
and replacing $\lambda^{(0)}$ by
$\lambda^{(0)}-\delta\lambda^{(0)}_T$ since this removes properly
the vacuum to vacuum diagrams at any order $m$ of perturbation
theory of $V^{\rm eff}_t$ and other diagrams give vanishing
contribution for $\phi=0$.

The (perturbative) thermodynamic limit $T_0\to \infty$ of $\rho_{T}^{\rm
ren}(\phi)$ can now be taken achieving at once the finiteness of
the perturbation expansion and the normalization of $\rho^{\rm
ren}(\phi)$ since ${\cal V}[G](T=0,T_0,\phi)$ converges as
$T_0\to\infty$ for $G\in \bar F_c(m)$ not a vacuum to vacuum diagram,
cf \cite[Section 7]{DGO}.

Of course, the above renormalization problem is rather trivial as the particle system only has short range $p$-body forces for $p\leq\bar p$. But having put the problem of TD limits of particles in the continuum in the language of the (generalized) renormalization group
now paves the way to the use of typical renormalization techniques, as e.g. differential inequalities and inductive proofs of renormalizability order by order in perturbation theory \cite{Sa}, to tackle less trivial problems in the thermodynamics of particle systems with long range forces.

\vspace{1cm}

\noindent {\sc Hanno Gottschalk, Boubaker Smii}\\
\rm Institut f\"ur angewandte Mathematik\\
Wegelerstr. 6\\
D-51373 Bonn, Germany\\
e-mails: gottscha/boubaker@wiener.iam.uni-bonn.de

\vspace{1cm}

\noindent {\sc Habib Ouerdiane}\\
 D\'epartement de Math\'ematique\\
Universit\'e de Tunis El Manar\\ Campus Universitaire, TN-1006
Tunis\\
e-mail: habib.ouerdiane@fst.rnu.tn
}
\end{document}